\documentclass[12 pt,twoside,aps,prd,amsmath,amssymb,
tightenlines,showpacs,showkeys,eqsecnum]{revtex4}

\usepackage{graphicx}
\usepackage{mathptmx}
\usepackage{bm}
\newcommand {\ve}{\varepsilon}

\newcommand {\bp}{\bar \psi}

\newcommand {\vf}{\varphi}

\def\myfigure #1#2#3#4
{\begin{figure}[ht]\begin{center}
\includegraphics[width=#2 \textwidth]{#1.eps}
\parbox[t]{#4\textwidth}{\caption{#3}\label{#1}}
\end{center}\end{figure}}

\begin{document}
\baselineskip -24pt
\title{Spinor model of a perfect fluid}
\author{Bijan Saha}
\affiliation{Laboratory of Information Technologies\\
Joint Institute for Nuclear Research, Dubna\\
141980 Dubna, Moscow region, Russia} \email{bijan@jinr.ru}
\homepage{http://wwwinfo.jinr.ru/~bijan/}

\begin{abstract}

Different characteristic of matter influencing the evolution of the
Universe has been simulated by means of a nonlinear spinor field. We
have considered two cases where the spinor field nonlinearity occurs
either as a result of self-action or due to the interaction with a
scalar field.

\end{abstract}



\maketitle

\bigskip

\section{Introduction}

In a series of papers we have studied the evolution of an
anisotropic universe where the source is given by a nonlinear spinor
field \cite{sahaprd,SBprd04,ECAA06,sahaprd06,BVI}. In those papers
it was shown that a suitable choice of nonlinearity (i) accelerates
the isotropization process, (ii) gives rise to a singularity-free
Universe and (iii) generates late time acceleration. In a recent
paper \cite{shikin} the authors have simulated perfect fluid using
spinor field with different nonlinearities. In doing so they used
the perfect fluid given by a barotropic equation of state. In this
paper beside the barotropic one we consider the Chaplygin gas as
well. As far as spinor field is concerned, we have considered two
possibilities: (i) the nonlinearity occurs as a result of self
action and (ii) the nonlinearity is a induced one, i.e., emerges due
to the interaction with a scalar field.

\section{Simulation of perfect fluid with nonlinear spinor field}

First of all let us note that one of the simplest and popular model
of the Universe is a homogeneous and isotropic one filled with a
perfect fluid with the energy density $\ve = T_0^0$ and pressure $p
= - T_1^1 = -T_2^2 = -T_3^3$ obeying the barotropic equation of
state
\begin{equation}
p = W \ve, \label{beos}
\end{equation}
where $W$ is a constant. Depending on the value of $W$ \eqref{beos}
describes perfect fluid from phantom to ekpyrotic matter, namely
\begin{subequations}
\label{zeta}
\begin{eqnarray}
W &=& 0, \qquad \qquad \qquad {\rm (dust)},\\
W &=& 1/3, \quad \qquad \qquad{\rm (radiation)},\\
W &\in& (1/3,\,1), \quad \qquad\,\,{\rm (hard\,\,Universe)},\\
W &=& 1, \quad \qquad \quad \qquad {\rm (stiff \,\,matter)},\\
W &\in& (-1/3,\,-1), \quad \,\,\,\,{\rm (quintessence)},\\
W &=& -1, \quad \qquad \quad \quad{\rm (cosmological\,\, constant)},\\
W &<& -1, \quad \qquad \quad \quad{\rm (phantom\,\, matter)},\\
W &>& 1, \quad \qquad \quad \qquad{\rm (ekpyrotic\,\, matter)}.
\end{eqnarray}
\end{subequations}
In order to describe the matter given by \eqref{zeta} with a spinor
field let us now write the corresponding Lagrangian \cite{sahaprd}:
\begin{equation}
L_{\rm sp} = \frac{i}{2} \biggl[\bp \gamma^{\mu} \nabla_{\mu} \psi-
\nabla_{\mu} \bar \psi \gamma^{\mu} \psi \biggr] - m\bp \psi + F,
\label{lspin}
\end{equation}
where the nonlinear term $F$ describes the self-interaction of a
spinor field and can be presented as some arbitrary functions of
invariants generated from the real bilinear forms of a spinor field.
For simplicity we consider the case when $F = F(S)$ with $S = \bp
\psi$. We consider the case when the spinor field depends on $t$
only. In this case for the components of energy-momentum tensor we
find
\begin{subequations}
\begin{eqnarray}
T_0^0 &=& mS - F, \label{t00s}\\
T_1^1 = T_2^2 = T_3^3 &=& S \frac{dF}{dS} - F. \label{t11s}
\end{eqnarray}
\end{subequations}

Inserting  $\ve = T_0^0$ and $p = - T_1^1$ into \eqref{beos} we find

\begin{equation}
S \frac{dF}{dS} - (1+W)F + m W S= 0, \label{eos1s}
\end{equation}
with the solution
\begin{equation}
F = \lambda S^{1+W} + mS, \label{sol1}
\end{equation}
with $\lambda$ being an integration constant. Inserting \eqref{sol1}
into \eqref{t00s} we find that
\begin{equation}
T_0^0 = - \lambda S^{1+W}. \label{lambda}
\end{equation}
Since energy density should be non-negative, we conclude that
$\lambda$ is a negative constant, i.e., $\lambda = - \nu$, with
$\nu$ being a positive constant. So finally we can write the
components of the energy momentum tensor
\begin{subequations}
\begin{eqnarray}
T_0^0 &=& \nu S^{1+W}, \label{t00sf}\\
T_1^1 = T_2^2 = T_3^3 &=& - \nu W S^{1+W}. \label{t11sf}
\end{eqnarray}
\end{subequations}
As one sees, the energy density $\ve = T_0^0$ is always positive,
while the pressure $p = - T_1^1 = \nu W S^{1+W}$ is positive for $W
> 0$, i.e., for usual fluid and negative for $W < 0$, i.e. for dark
energy.

In account of it the spinor field Lagrangian now reads
\begin{equation}
L_{\rm sp} = \frac{i}{2} \biggl[\bp \gamma^{\mu} \nabla_{\mu} \psi-
\nabla_{\mu} \bar \psi \gamma^{\mu} \psi \biggr] - \nu S^{1+W},
\label{lspin1}
\end{equation}
Thus a massless spinor field with the Lagrangian \eqref{lspin1}
describes perfect fluid from phantom to ekpyrotic matter. Here the
constant of integration $\nu$ can be viewed as constant of
self-coupling. A detailed analysis of this study was given in
\cite{shikin}.

Let us now generate a Chaplygin gas by means of a spinor field. A
Chaplygin gas is usually described by a equation of state
\begin{equation}
p = -A/\ve^\gamma. \label{chap}
\end{equation}
Let us consider the case with $\gamma = 1$. Then in case of a
massless spinor field for $F$ one finds
\begin{equation}
\frac{F dF}{F^2 - A} = 2 \frac{dS}{S}, \label{eqq}
\end{equation}
with the solution
\begin{equation}
F = \pm \sqrt{A + \lambda S^2}. \label{chapsp}
\end{equation}
Since, in this case, $T_0^0 = -F$ should be nonnegative, the
expression for $F$ should be negative. On account of this for the
components of energy momentum tensor we find
\begin{subequations}
\begin{eqnarray}
T_0^0 &=& \sqrt{A + \lambda S^2}, \label{edchapsp}\\
T_1^1 = T_2^2 = T_3^3 &=& A/\sqrt{A + \lambda S^2}. \label{prchapsp}
\end{eqnarray}
\end{subequations}
AS was expected, we again get positive energy density and negative
pressure.

Thus the spinor field Lagrangian corresponding to a Chaplygin gas
reads
\begin{equation}
L_{\rm sp} = \frac{i}{2} \biggl[\bp \gamma^{\mu} \nabla_{\mu} \psi-
\nabla_{\mu} \bar \psi \gamma^{\mu} \psi \biggr] - \sqrt{A + \lambda
S^2}. \label{lspin2}
\end{equation}
Thus we see that a nonlinear spinor field with specific type of
nonlinearity can substitute perfect fluid and dark energy, thus give
rise to a variety of evolution scenario of the Universe.

\section{Simulation of perfect fluid with interacting spinor and scalar fields}

Now let us consider the system with interacting spinor and scalar
fields with the Lagrangian \cite{SBprd04}:
\begin{equation}
L_{\rm int} = \frac{i}{2} \biggl[\bp \gamma^{\mu} \nabla_{\mu} \psi-
\nabla_{\mu} \bar \psi \gamma^{\mu} \psi \biggr] - m \bp \psi +
\frac{1}{2} \vf_{,\alpha} \vf^{,\alpha} (1 + \lambda_1 F_1),
\label{lspsc}
\end{equation}
where $F_1= F_1(S)$.  We again consider the case with spinor and
scalar fields being the functions of $t$ only. Denoting $F_2 = 1 +
\lambda_1 F_1$ in this case for the components of energy-momentum
tensor we find
\begin{subequations}
\begin{eqnarray}
T_0^0 &=& mS +\frac{1}{2}\frac{S^2}{F_2}, \label{t00sc}\\
T_1^1 = T_2^2 = T_3^3 &=& \frac{1}{2}\frac{S^2}{F_2^2}(S
\frac{dF_2}{dS} - F_2). \label{t11sc}
\end{eqnarray}
\end{subequations} Here we have taken into account that ${\dot \vf}^2
= Q S^2/F_2^2$ \cite{bviss}. Here $Q$ is a constant that depends on
the concrete cosmological model. Here we also used the fact that $S
= C_0/\tau$ with $\tau = abc$. For simplicity we set $Q = 1$ and
$C_0 = 1$. These equalities hold for the cosmological models given
by \cite{BVI}
\begin{equation}
ds^2 = dt^2 - a^{2} e^{-2mz}\,dx^{2} - b^{2} e^{2nz}\,dy^{2} -
c^{2}\,dz^2. \label{bvi}
\end{equation}
Here $a,b,c$ are the functions of $t$ only and depending on the
value of $m, n$ \eqref{bvi} describes Bianchi type VI, V, III, I and
FRW cosmological models \cite{BVI}.

Inserting \eqref{t00sc} and \eqref{t11sc} into \eqref{beos} we find
\begin{equation}
S^2 \frac{dF_2}{dS} + (W-1) S F_2 + 2 mW F_2^2 = 0. \label{eosint1}
\end{equation}
From \eqref{eosint1} one finds
\begin{equation}
F_2 = -\frac{1}{2m} S. \label{sol2}
\end{equation}
As one sees, the spinor field nonlinearity in this case does not
depend on $W$. Moreover, inserting this into \eqref{t00sc} one finds
$T_0^0 = -mS$. So we can dully neglect this case and move forward
with a massless spinor field. In this case from \eqref{eosint1} we
find
\begin{equation}
F_2 = C S^{1-W}. \label{sol3}
\end{equation}
In this case for the components of the energy momentum tensor we
find
\begin{subequations}
\begin{eqnarray}
T_0^0 &=& \frac{1}{2C} S^{1+W}, \label{t00scn}\\
T_1^1 = T_2^2 = T_3^3 &=& -\frac{W}{2C} S^{1+W}, \label{t11scn}
\end{eqnarray}
\end{subequations}
which, as was expected coincides with the one given by \eqref{t00sf}
and \eqref{t11sf} with $\nu = 1/2C$.

As far as Chaplygin gas concerned, inserting \eqref{t00sc} and
\eqref{t11sc} into \eqref{chap} for massless spinor field we find
\begin{equation}
S \frac{dF_2}{dS} - 4 A \frac{F_2^3}{S^4} - F_2 = 0, \label{eosint2}
\end{equation}
with the solution
\begin{equation}
F_2 = \frac{1}{2\sqrt{A}} S^2. \label{sol4}
\end{equation}
In this case for the components of the energy momentum tensor we
find
\begin{subequations}
\begin{eqnarray}
T_0^0 &=& \sqrt{A}, \label{t00scnch}\\
T_1^1 = T_2^2 = T_3^3 &=& \sqrt{A}, \label{t11scnch}
\end{eqnarray}
\end{subequations}
which coincides with \eqref{edchapsp} and \eqref{prchapsp} when
$\lambda = 0$.

Thus we see that an interacting system of spinor and scalar field
can as well describe a perfect fluid and dark energy.

\section{Anisotropic cosmological models with a spinor field}

In the previous two sections we showed that the perfect fluid and
the dark energy can be simulated by a nonlinear spinor field. In the
section II the nonlinearity was the subject to self-action, while in
section III the nonlinearity was induced by a scalar field. It was
also shown the in our context the results of section III is some
special cases those of section II. Taking it into mind we study the
evolution an anisotropic Universe filled with a nonlinear spinor
field given by the Lagrangian \eqref{lspin}, with the nonlinear term
$F$ is given by \eqref{sol1} of \eqref{chapsp}.

\section{Conclusion}

Within the framework of cosmological gravitational field equivalence
between the perfect fluid (and dark energy) and nonlinear spinor
field has been established. It is shown that the perfect fluid can
be simulated with both self-action and induced nonlinearity of the
spinor field. The case with induced nonlinearity can be viewed as
partial case that of self-action.

\end{document}